\documentclass[Journal,letterpaper]{ascelike-new}
\usepackage[utf8]{inputenc}
\usepackage[T1]{fontenc}

\usepackage{fullpage}
\usepackage{subcaption}
\usepackage{lmodern}
\usepackage{graphicx}
\usepackage[figurename=Fig.,labelfont=bf,labelsep=period]{caption}
\usepackage{natbib}
\usepackage{subcaption}
\usepackage{amsmath}
\usepackage{newtxtext,newtxmath}
\usepackage[colorlinks=true,citecolor=red,linkcolor=black]{hyperref}

\usepackage[scr=boondoxo]{mathalfa}

\newcommand{\G}{\mathcal{G}}
\newcommand{\N}{\mathcal{N}}
\newcommand{\Z}{\mathcal{Z}}
\newcommand{\A}{\mathcal{A}}
\newcommand{\C}{\mathcal{C}}
\newcommand{\X}{\mathcal{X}}
\NameTag{Cotta Antunez \& Levin, \today}
\begin{document}
\nolinenumbers
\title{How Does Eco-Routing Affect Total System Emissions? City Network Predictions From User Equilibrium Models}

\author[1]{Roc\'{\i}o Cotta Ant\'unez}
\author[2]{Michael W. Levin}

\affil[1]{Department of Computer Science and Engineering, University of Minnesota. Email: cotta033@umn.edu}
\affil[2]{Department of Civil, Environmental, and Geo- Engineering, University of Minnesota. Email: mlevin@umn.edu}

\maketitle

\begin{abstract}
Transportation contributes a substantial fraction of all greenhouse gas emissions. One approach for reducing such emissions is to modify vehicles' route choices to minimize their fuel consumption or emission, which is known as eco-routing. 
Most eco-routing is based on vehicles choosing routes that minimize their individual fuel consumption or emissions. 
The Braess paradox demonstrates that when vehicles choose routes to minimize their individual goals, the aggregate effect may paradoxically result in the opposite net effect due to changes in congestion patterns.
We construct a multiclass user equilibrium model in which some vehicles use eco-routing and others seek to minimize their individual travel times. Using this model, we show that the Braess paradox exists for eco-routing. If a large number of vehicles are trying to minimize their fuel consumption or emissions, the total fuel consumption or emissions may increase. We then solve the multiclass user equilibrium on publicly available city network data, and find that eco-routing results in increases in fuel consumption and emissions on some city networks as well.
\end{abstract}

\section{Introduction}

Transportation contributes a substantial fraction of all greenhouse gas emissions, and reducing the fuel consumption and emissions from transportation is therefore beneficial. 
There are multiple approaches  for reducing the emissions from transportation, including more efficient engine technologies, alternative fuels, reduction in vehicle miles traveled, etc. 
Fuel consumption and emissions also vary with driving characteristics, such as acceleration, speed, and road grade~\citep{franco2013road}. Since driving characteristics vary with individual roads and time-of-day, route choices can affect the fuel consumption and emissions from a given origin-destination vehicle trip. Previous work has proposed adjusting route choices to reduce fuel consumption and/or emissions~\citep{ahn2013network}, and we refer to such route choice behavior as \textit{eco-routing}. Drivers who choose routes to minimize their own travel time, in contrast, are referred to as \textit{time-routing}.

Transportation models have typically assumed that drivers seek to minimize their own travel time~\citep{wardrop1952road}, and  smartphone navigation apps provide real-time routing guidance for users to minimize travel times.
Since drivers try to minimize their own travel times and not the total congestion in the system, the behavior results in an user equilibrium of route choices in which no driver can improve their travel time by changing routes. 
Although eco-routing has previously been discussed in the literature~\citep{zhou2016review}, 
emissions and fuel consumption per road are not as easy for drivers to estimate, which has made eco-routing difficult to implement in practice.
However, in 2021 Google Maps introduced eco-routing in the United States with the stated goal of guiding drivers on routes with the ``lowest carbon footprint''~\citep{GoogleMaps}. Moreover, the default mode for Google Maps directions is now eco-routing, so users have to modify their settings to avoid it. Consequently, this change may have already resulted in  a large percentage of drivers using eco-routing.

The goal of such eco-routing is to reduce carbon emissions from transportation.
We define \textit{total system emissions} (TSE) as the sum of the carbon emissions over all vehicles traveling through the system.
An individual driver using eco-routing will likely achieve reductions in emissions and/or fuel consumption for themselves, corresponding to tiny impacts in TSE. However, if a large number of users switch to eco-routing, the impacts on TSE are unclear because large-scale changes in route choice affect traffic congestion, which affects emissions. A similar problem is well-known to occur with time-routing. When all drivers choose routes to minimize their own travel time, the resulting user equilibrium is often substantially worse than the minimum total system travel time that can be achieved~\citep{roughgarden2003price}. \citet{braess1968paradoxon} demonstrated paradoxically that user equilibrium can cause improvements to network infrastructure to actually increase traffic congestion and the total system travel time due to changes in route choices. 
We predict that a large-scale shift to eco-routing could result in a similar paradox and increase TSE.
Although \citet{ahn2013network} found that eco-routing would reduce emissions in Cleveland and Columbus, Ohio, that does not guarantee reductions in emissions in all cities.
The purpose of this paper is to demonstrate this paradox both in small intuitive examples and in large, realistic networks. 
We hope that this demonstration will encourage network analyses of eco-routing to determine its effectiveness in specific cities prior to  implementing it in practice.

The contributions of this paper are as follows: we define and solve a multiclass static traffic assignment problem where some users choose eco-routing and others choose time-routing. Then, we study a simple 2-link network to build intuition on when eco-routing is likely to increase total system emissions. We also solve traffic assignment on city networks to demonstrate that eco-routing could increase total system emissions for the entire city.
These numerical results appear to be the first demonstration in the literature that eco-routing could potentially increase emissions due to the congestion changes from new routing behaviors.

\section{Literature review}

\subsection{Models}
Eco-routing can be used to find the paths that will minimize the fuel consumption or carbon emissions for an individual vehicle. It can be implemented using different types of algorithms, both macroscopic \citep{dhaou2011fuel}, such as the ant colony optimization \citep{elbery2016eco}, and microscopic. \citet{kubivcka2016prediction} presented a microscopic standard model that is reformulated into a macroscopic model to predict consumption. Some other models with a microscopic approach focus on the characteristics of the links in a network \citep{6060054}, location-based attributes \citep{minett2011eco}, as well as the vehicles' operating conditions \citep{nie2013eco} and their trajectories \citep{sun2015stochastic}. Furthermore, examples of microscopic algorithms that have been used in eco-routing models can be based on individual agents or sub-populations \citep{rakha2012integration}. \citet{levin2014effect} observed that road grade could have a large effect on the performance of eco-routing algorithms.

One of the applications of eco-routing includes creating navigation systems that will find the lowest energy consumption routes for fuel and electric vehicles \citep{wang2019real}. In \citet{das2019road}, the analysis for navigation systems for electric vehicles takes into account the road load to create the its model. There exists eco-routing models for each particular type of vehicle, since there are differences on the implementation that will get the most energy efficient route for vehicles with combustion engines than for hybrid electric vehicles (battery powered and plug-in) \citep{richter2012comparison}. For hybrid plug-in vehicles, the route selection has to consider other factors like power-train control to minimize the energy consumption.

Power management has been included into the route selection process in different ways. One way is to simultaneously calculate the energy-optimal route considering the paths and power-train strategies. The results collected from running this algorithm in the SUMO traffic simulator showed significant energy savings for Boston \citep{houshmand2021combined} and Ann Harbor \citep{li2020eco}. As showed in \citet{houshmand2018eco}, this approach outperforms the traditional charge depleting first (CDF) models that had a fixed power-train control strategy. In \citet{caspari2021optimal}, they propose an optimization problem between the combustion engine and electric motor. Other factors that present challenges when finding the most energy efficient routes are limited on-board energy storage and the effect of traffic on energy consumption and travel time \citep{guanetti2019eco}.

While most eco-routing models focus on minimizing a single factor, whether it is energy or fuel consumption, there exists multi-objective models that cover more than one factor, such as both energy consumption and travel time \citep{ahn2021multi} or travel time, vehicle kilometres travelled, greenhouse gas, and Nitrogen Oxide \citep{alfaseeh2019multi}. Since heavy trucks consume large amounts of fuel and emit a lot of carbon emissions, \citet{scora2015value} presents an eco-routing model specified in reducing energy and emission consumption for these vehicles. In the same way, \citet{lu2016eco} creates a model to find the flows that minimize emissions in congested networks. Some eco-routing models have an extra constraint besides minimizing carbon emissions: a travel time budget. In \citet{zeng2020eco}, the goal of including the time constraint is to allow navigation systems to plan a trip and ensure arrival at a specific time. Other models add the travel time limitation to act as a trade-off to reduce carbon emissions by using a k-shortest path algorithm \citep{zeng2016prediction,zeng2017application}.

\subsection{Effects}
Eco-routing is implemented to reduce energy consumption. In order to check if these models are having the desired effects on real-life networks, studies are conducted to analyze eco-routing in several cities. In \citet{bandeira2012comparative}, data collected from Hampton Roads, VA and the city of Aveiro, Portugal shows that implementing eco-routing reduced both global and local pollutants during peak-hours and off-peak at the same rate. The effects of eco-routing in the small city of Caceres, Spain were satisfactory in urban roads where carbon emissions decreased. However, in bypasses, the eco-routing model resulted in an increase of fuel consumption and emissions \citep{coloma2019environmental}. Moreover, implementing an algorithm based on neural networks, data from the Spanish transmission system operator, eco-routing, eco-driving and eco-charging in the city of Alcalá de Henares in Madrid, Spain contributed to daily energy savings \citep{ortega2021can}. Finally, a study on eco-routing systems on the large-scale networks of the cities of Columbus and Cleveland in Ohio demonstrated that the fuel savings were achieved, which were sensitive to the network configuration and level of market penetration of the eco-routing system \citep{ahn2013network}.
Overall, most studies expect that large-scale eco-routing will result in environmental benefits. To the best of our knowledge, no study has yet demonstrated how eco-routing could increase fuel consumption and/or emissions due to the changes it causes in congestion patterns.

\section{Methods}

In this section, we define the multiclass static traffic assignment problem that we will use in the numerical results. Consider a network $\G=(\N,\A)$ where $\N$ is the set of nodes and $\A$ is the set of links. 
Let $t_{ij}(x_{ij})$, $f_{ij}(x_{ij})$, and $e_{ij}(x_{ij})$ be the travel time, fuel consumption, and energy consumption, respectively, for link $(i,j)$ when the flow on link $(i,j)$ is $x_{ij}$. 
Let $\Z\subseteq\N$ be the set of zones. We consider two classes of vehicles. Let  $\C=\{\mathrm{t}, \mathrm{e}, \mathrm{f}\}$ be the set of classes where $\mathrm{t}$, $\mathrm{e}$, and $\mathrm{f}$ denote drivers seeking to minimize their travel time, emissions, and fuel consumption, respectively.
Let $d^c_{rs}$ be the demand of class $c$ from zone $r$ to zone $s$.

\subsection{User equilibrium}

Consider a path $\pi$ and let $h^c_\pi$ be the flow on path $\pi$ of class $c$. 
Let $E_\pi=\sum_{(i,j)\in\pi} e_{ij}(x_{ij})$, $F_\pi=\sum_{(i,j)\in\pi} f_{ij}(x_{ij})$, and $T_\pi=\sum_{(i,j)\in\pi} t_{ij}(x_{ij})$ be the emissions, fuel consumption, and travel time of path $\pi$, respectively.
Let $\Pi$ be the set of all paths, and let $\Pi_{rs}\subseteq\Pi$ be the set of paths with origin $r$ and destination $s$.
Eco-routing vehicles seek to minimize the emissions or fuel consumption of their path. Equivalently, a path $\pi$ from $r$ to $s$ is used only if it achieves the minimum cost $\mu^c_{rs}$ for travel from $r$ to $s$, which can be written as 
\begin{linenomath*}
\begin{subequations}
\label{1}
\begin{align}
    h^{\mathrm{e}}_\pi \left(E_\pi - \mu^{\mathrm{e}}_{rs} \right) &= 0 && \forall \pi\in\Pi \label{1a}\\
    h^{\mathrm{f}}_\pi \left(F_\pi - \mu^{\mathrm{f}}_{rs} \right) &= 0 && \forall \pi\in\Pi \label{1b}\\
    h^{\mathrm{t}}_\pi \left(T_\pi - \mu^{\mathrm{t}}_{rs} \right) &= 0 && \forall \pi\in\Pi  \label{1c}
\end{align}
\end{subequations}
\end{linenomath*}
where conditions \eqref{1a}, \eqref{1b}, and \eqref{1c} must hold for vehicles seeking to minimize travel time, emissions, and fuel consumption, respectively.

\subsection{Cost functions}

We use the Bureau of Public Roads function for travel times:
\begin{linenomath*}
\begin{align}
    t_{ij}(x_{ij})=t^{\mathrm{ff}}_{ij} \left(1 + \alpha_{ij} \left( \frac{x_{ij}}{Q_{ij}} \right)^\beta \right)
\end{align}
\end{linenomath*}
where $t^{\mathrm{ff}}_{ij}$ is the free flow travel time, $Q_{ij}$ is the link capacity, and $\alpha_{ij}$ and $\beta_{ij}$ are calibration constants for link $(i,j)$.
For emissions and fuel consumption, we use the static traffic assignment functions developed by \citet{gardner2013framework} for internal combustion engine vehicles. These functions were developed through regression on data from the US Environmental Protection Agency (EPA)'s MOVES software~\citep{vallamsundar2011overview}. 
\citet{gardner2013framework} also developed functions for plug-in electric vehicles, so this analysis method could be extended to electric vehicles.
For fuel consumption,
\begin{linenomath*}
\begin{align}
    f_{ij}(x_{ij}) = \ell_{ij} \times 14.58 \left(u_{ij}(x_{ij}) \right)^{-0.6253}
\end{align}
\end{linenomath*}
where $f_{ij}(x_{ij})$ is energy consumption in kWh, $\ell_{ij}$ is the length of link $(i,j)$ in miles, and $u_{ij}(x_{ij})$ is the speed on link $(i,j)$ in mi/hr, defined as 
\begin{linenomath*}
\begin{align}
    u_{ij}(x_{ij}) = \frac{\ell_{ij}}{t_{ij}(x_{ij})} \label{fuel}
\end{align}
\end{linenomath*}
The free flow time $t^{\mathrm{ff}}_{ij}$ is related to link length via  $t^{\mathrm{ff}}_{ij} = \ell_{ij}/ u^{\mathrm{ff}}_{ij}$ where $u^{\mathrm{ff}}_{ij}$ is the free flow speed.
Although $f_{ij}(x_{ij})$ is specified in kWh, it can be converted to gallons of fuel if the conversion factor is known.
\citet{gardner2013framework} defined functions for the emissions of carbon dioxide, nitrogen oxides, and volatile organic compounds. Since the stated goal of Google Maps is to reduce carbon emissions~\citep{GoogleMaps}, we define our emissions functions in terms of carbon dioxide:
\begin{linenomath*}
\begin{align}
    e_{ij}(x_{ij}) = \ell_{ij} \times 3158 \left(u_{ij}(x_{ij})\right)^{-0.56} \label{co2}
\end{align}
\end{linenomath*}
where $e_{ij}(x_{ij})$ is in grams.
Because equations \eqref{fuel} and \eqref{co2} are monotone decreasing with respect to $u_{ij}(x_{ij})$, they are monotone increasing with respect to $t_{ij}(x_{ij})$. Since we use the BPR function for $t_{ij}(x_{ij})$, which is  monotone increasing with respect to $x_{ij}$, equations \eqref{fuel} and \eqref{co2} are also monotone increasing with respect to $x_{ij}$.

\subsection{Traffic assignment problem}

The traffic assignment problem is to find link flows that satisfy user equilibrium conditions \eqref{1}. We must first define the feasible set of link flows, $\X$. To do this, we disaggregate the link flow $x_{ij}$ into class-specific link flows $x^c_{ij}$, where $x_{ij}=\sum_{c\in\C} x^c_{ij}$. Any $x^c_{ij}\in\X$ must be determined by path flows $h^c_\pi$ via
\begin{linenomath*}
\begin{align}
    x^c_{ij} =\sum\limits_{\pi\in\Pi}  \delta^\pi_{ij} h^c_{\pi} && \forall (i,j)\in\A, \forall c\in\C \label{eq6}
\end{align}
\end{linenomath*}
Path flows must correspond to the network demand:
\begin{linenomath*}
\begin{align}
    d^c_{rs} = \sum\limits_{\pi\in\Pi_{rs}} h^c_\pi && \forall (r,s)\in\Z^2, \forall c\in\C
\end{align}
\end{linenomath*}
and must be non-negative:
\begin{linenomath*}
\begin{align}
    h^c_\pi \geq 0 && \forall \pi\in\Pi, \forall c\in\C \label{eq8}
\end{align}
\end{linenomath*}
Together, equations \eqref{eq6}--\eqref{eq8} define the set of feasible link flows $\X$. Our goal is to find a link flow assignment in $\X$ that also satisfies user equilibrium. 
\citet{dial1999network} formulated a multiclass traffic assignment problem with tolls as a variational inequality (VI). We adopt a similar approach to combine time-routing and eco-routing vehicles here: find $( x^{\mathrm{t}\star}_{ij}, x^{\mathrm{e}\star}_{ij}, x^{\mathrm{f}\star}_{ij})\in \X$ such that for all $( x^{\mathrm{t}}_{ij}, x^{\mathrm{e}}_{ij}, x^{\mathrm{f}}_{ij})\in \X$, 
\begin{linenomath*}
\begin{align}
    \sum\limits_{(i,j)\in\A} t_{ij}\left(x^{\star}_{ij}\right) \left(x^{\mathrm{t}}_{ij} - x^{\mathrm{t}\star}_{ij}\right)
    + \sum\limits_{(i,j)\in\A} e_{ij}\left(x^{\star}_{ij}\right) \left(x^{\mathrm{e}}_{ij} - x^{\mathrm{e}\star}_{ij}\right)
    + \sum\limits_{(i,j)\in\A} f_{ij}\left(x^{\star}_{ij}\right) \left(x^{\mathrm{f}}_{ij} - x^{\mathrm{f}\star}_{ij}\right)
    \geq 0
\end{align}
\end{linenomath*}
When all demand is of a single class (including 100\% eco-routing vehicles), this VI is convex because $t_{ij}(x_{ij})$, $e_{ij}(x_{ij})$, and $f_{ij}(x_{ij})$ are all monotone increasing with respect to $x_{ij}$. 
Among other things, convexity means that the solution is unique. 
However, when multiple classes are present, convexity is not guaranteed~\citep{marcotte2004new} and consequently there may not be an unique solution.
We use the method of successive averages to find an equilibrium, and validate it by evaluating the equilibrium gap, $g$:
\begin{linenomath*}
\begin{align}
    g=\frac{
    \sum\limits_{(i,j)\in\A} \left[ e_{ij}(x_{ij}) x^{\mathrm{e}}_{ij} + f_{ij}(x_{ij}) x^{\mathrm{f}}_{ij} + t_{ij}(x_{ij}) x^{\mathrm{t}}_{ij}\right] 
     - \sum\limits_{(r,s)\in\Z^2} \sum\limits_{c\in\C} d^c_{rs} \mu^c_{rs}
    }{
\sum\limits_{(i,j)\in\A} \left[ e_{ij}(x_{ij}) x^{\mathrm{e}}_{ij} + f_{ij}(x_{ij}) x^{\mathrm{f}}_{ij} + t_{ij}(x_{ij}) x^{\mathrm{t}}_{ij}\right] 
    }
\end{align}
\end{linenomath*}
The gap is the percent difference between the actual travel cost and the minimum travel cost. 
When the gap is 0, the equilibrium conditions \eqref{1} are satisfied exactly. In practice, we determine that we reached the equilibrium when the gap is below a certain predefined threshold that we will call $\epsilon$.

\section{Numerical results}

Using the traffic assignment methods previously described, we now study the impacts of eco-routing user equilibrium on the overall network.
We are primarily interested in comparing two metrics: total system emissions (TSE), and total system fuel consumption (TSFC). These are defined as follows:
\begin{linenomath*}
\begin{subequations}
\begin{align}
        \mathrm{TSE} &= \sum\limits_{(i,j)\in\A} e_{ij}(x_{ij}) \times x_{ij}
        \\
            \mathrm{TSFC} &= \sum\limits_{(i,j)\in\A} f_{ij}(x_{ij}) \times x_{ij}
\end{align}
\end{subequations}
\end{linenomath*}
We aim to demonstrate that a shift from time-routing to eco-routing can cause TSE and/or TSFC to increase. In other words, a large number of vehicles seeking to minimize their individual emissions can cause their collective emissions to increase due to changing congestion patterns. 
We will demonstrate this increase in two ways. First, we conduct experiments on a simple 2-link network which provides an easily verifiable model where the causes of paradoxical results can be easily understood. Then, we extend our results to city networks to explore the possibility of such behavior in reality.

\subsection{2-link network}

\begin{figure}
    \centering
    \includegraphics[width=0.5\textwidth]{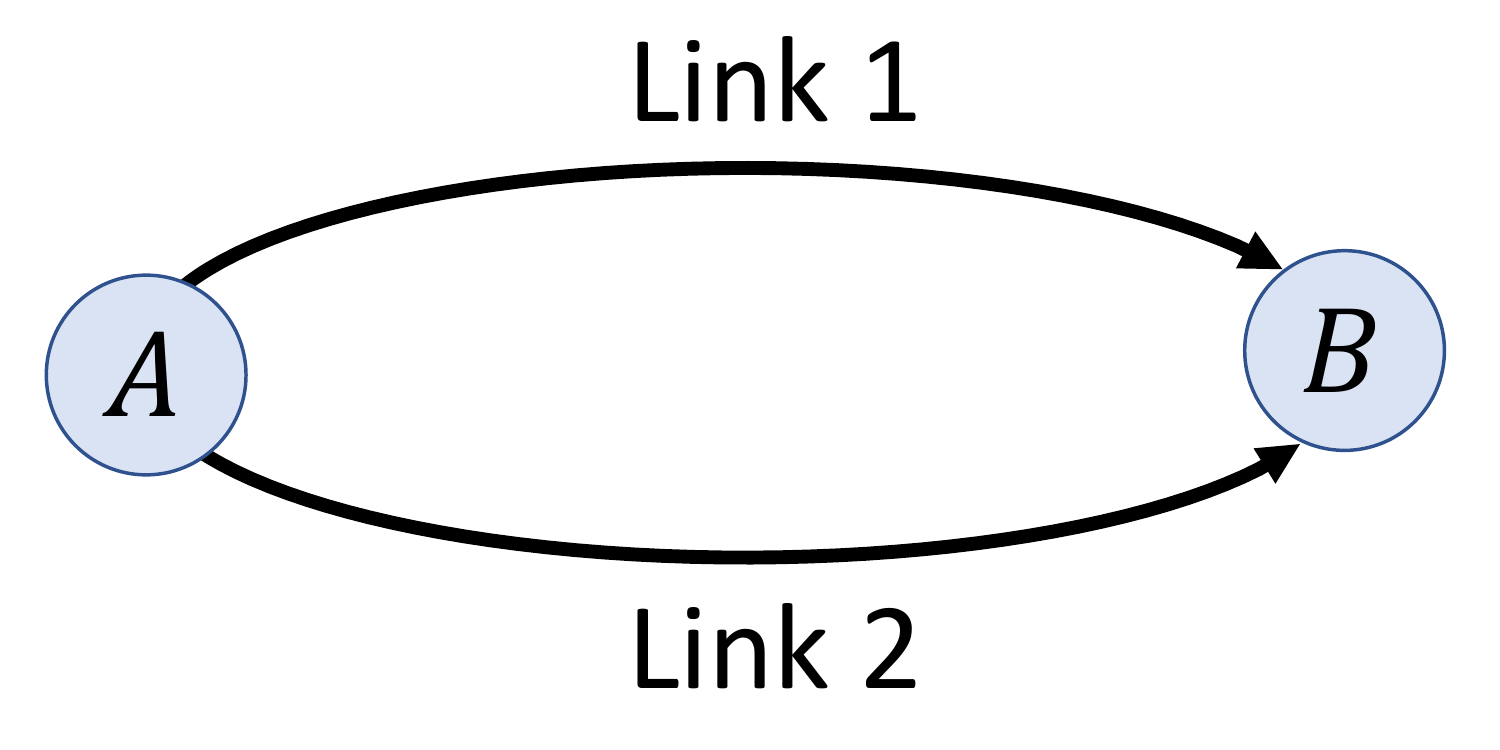}
    \caption{2-link network used to illustrate equilibrium impacts}
    \label{2link}
\end{figure}

We use the 2-link network shown in Figure \ref{2link} to explore how changing link parameters will affect fuel consumption and emissions from eco-routing. Link 1 has capacity $Q_1=1000$veh/hr, length $\ell_1=10$mi, free flow speed $u^{\mathrm{ff}}_1=45$mi/hr, and calibration parameters $\alpha_1=0.15$ and $\beta_1=4$. Link 2 has capacity $Q_2=2000$veh/hr and calibration parameters $\alpha_2=0.15$ and $\beta_2=4$, but the length $\ell_2$ and free flow speed $u^{\mathrm{ff}}_2$ are varied to explore the effects on the user equilibrium. The demand is 4000vph from $A$ to $B$, and all demand must use either link 1 or link 2.
We consider scenarios in which 100\% of vehicles are time-routing or 100\% of vehicles are eco-routing. 
We do not consider mixtures of time-routing and eco-routing demand in these 2-link network experiments to avoid multiple equilibria. 

We first demonstrate parameters for link 2 that cause eco-routing to increase TSFC and TSE. We choose $\ell_2=5$mi and $u^{\mathrm{ff}}_2=30$mi/hr, so link 2 is shorter than link 1, but has a lower free flow speed. 
These parameters could possibly describe a realistic road.
Link 2 also has a higher capacity, which could result from having more lanes than link 1. We compare three equilibrium scenarios: 100\% time-routing vehicles, 100\% eco-routing vehicles seeking to minimize their fuel consumption, and 100\% eco-routing vehicles seeking to minimize their CO2 emissions. Table \ref{2linkdemo} reports the travel time, fuel consumption, and emissions per vehicle for both links 1 and 2, and the total for the system. We can verify that equilibrium was found by comparing the link 1 and link 2 parameters. For 100\% time-routing vehicles, $t_1(x_1)=t_2(x_2)$. 
For 100\% eco-routing vehicles to minimize emissions, $e_1(x_1)=e_2(x_2)$.
For 100\% eco-routing vehicles to minimize fuel consumption, $f_1(x_1)=f_2(x_2)$. These satisfy the user equilibrium conditions \eqref{1}, showing that no vehicle can improve the travel cost that they care about by changing routes.

The results in Table \ref{2linkdemo} show that changing routing behavior from time-routing to eco-routing increased TSE by 10.2\% for eco-routing vehicles minimizing emissions and 5.1\% for eco-routing vehicles minimizing fuel consumption. These increases are significant considering the fact that eco-routing may  be implemented with the goal of decreasing the total emissions from transportation~\citep{GoogleMaps}. Instead, we might observe eco-routing causing emissions to increase.
Similar increases were observed in fuel consumption; TSFC increased by 13.5\% and 7.4\% for eco-routing to minimize emissions and fuel consumption, respectively. 
These results suggest that for this specific example, eco-routing to minimize fuel consumption had less of an adverse impact than eco-routing to minimize emissions. However, both were worse than vehicles using time-routing.

These results appear to occur because emissions and fuel consumption increase with $x_{ij}$ at a different rate than travel times. When eco-routing is used, the reported flows on link 2 have to be higher than when time-routing is used to achieve equilibrium.  Specifically, when time-routing is used, link 2 has a flow of 2881.5 vph and emissions of 3107.7 g/veh. When vehicles eco-route to minimize emissions, link 2's flow increases to 3451.2 vph so that both links 1 and 2 will have emissions of 3774.9 g/veh. This results in a larger number of vehicles using link 2 while link 2's emissions also increase to match link 1. Eco-routing achieves a reduction in emissions per vehicle on the lesser-used link 1, but that results in a higher TSE.

\begin{table}
\caption{Effects of eco-routing on travel time, fuel consumption, and emissions for the 2-link network}
    \label{2linkdemo}
    
    \centering
    \begin{tabular}{l|r|r|r|r}
    \hline
          &  Link flow (vph) & Travel time (min) & Fuel consumption (kWh) & Emissions (g CO2)\\\hline
         \multicolumn{5}{c}{100\% time-routing} \\\hline
         Link 1 & 1118.5 & 16.5 & 15.4 & 4215.9\\ 
         Link 2 & 2881.5 & 16.5 & 11.9 & 3107.7\\
         Total & 4000 & 65853.5 & 51419.9 & 1.37E7\\
         \hline
                  \multicolumn{5}{c}{100\% eco-routing, minimize CO2 emissions}\\\hline
         Link 1 & 548.8 & 13.5 & 13.6 & 3774.9\\
         Link 2 & 3451.2 & 23.3 & 14.7 & 3774.9\\
         Total & 4000 & 87828.1 & 58370.5 & 1.51E7\\
         \hline
         \multicolumn{5}{c}{100\% eco-routing, minimize fuel consumption}\\\hline
         Link 1 & 710.8 & 13.8 & 13.8 & 3826.1\\
         Link 2 & 3289.2 & 21.0 & 13.8 & 3559.0\\
         Total & 4000 & 78824.4 & 55242.8 & 1.44E7\\
         \hline

    \end{tabular}
    
\end{table}

\begin{figure}
    
    \begin{center}
    \begin{subfigure}{0.75\textwidth}
        \includegraphics[width=\textwidth]{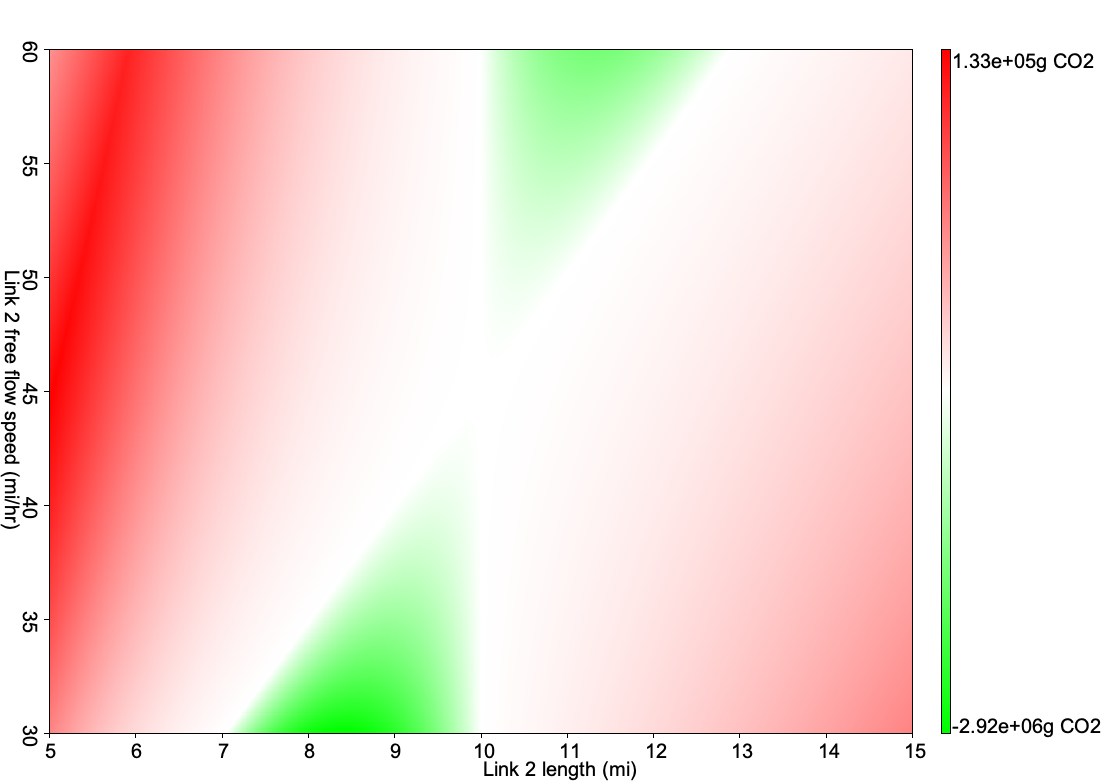}
        \caption{CO2 emissions: $\mathrm{TSE}_{\mathrm{eco-routing}}-\mathrm{TSE}_{\mathrm{time-routing}}$}
    \end{subfigure}

    \begin{subfigure}{0.75\textwidth}
        \includegraphics[width=\textwidth]{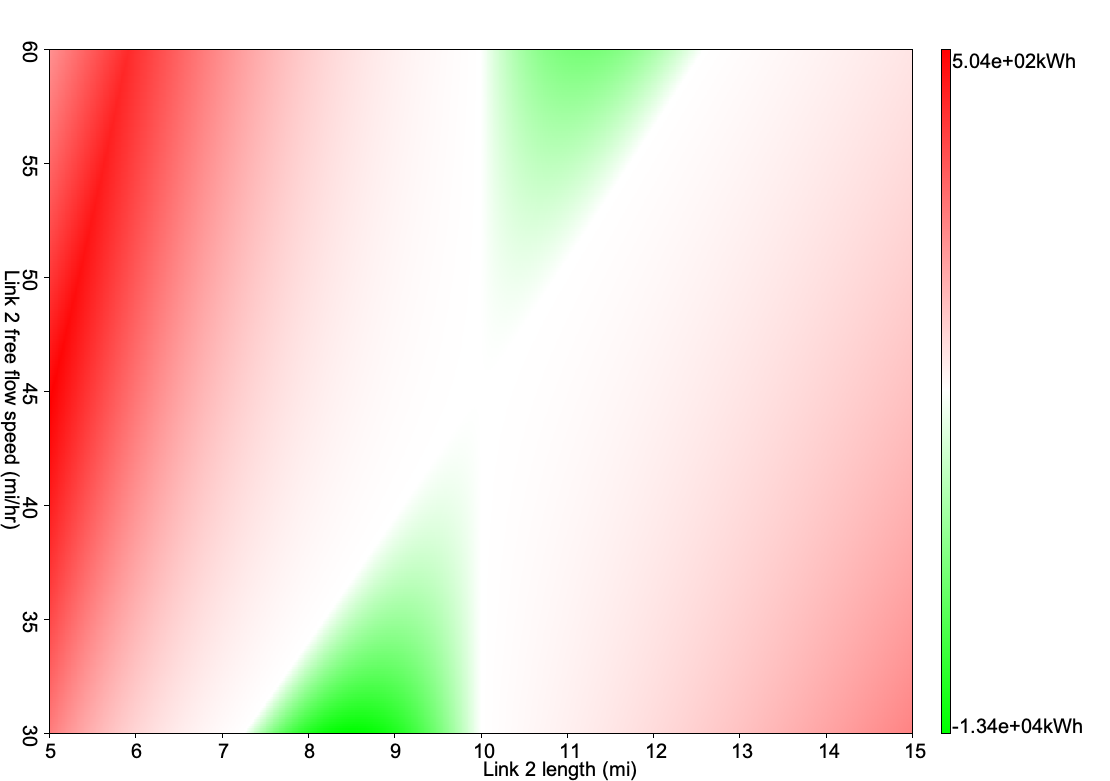}
        \caption{Fuel consumption: $\mathrm{TSFC}_{\mathrm{eco-routing}}-\mathrm{TSFC}_{\mathrm{time-routing}}$}
    \end{subfigure}
    \end{center}
    
    \caption{Change in emissions and fuel consumption caused by eco-routing to minimize emissions with respect to changing the link length and free flow speed of link 2.}
    \label{2linkheatmap_emissions}
\end{figure}

\begin{figure}
    
    \begin{center}
    \begin{subfigure}{0.75\textwidth}
        \includegraphics[width=\textwidth]{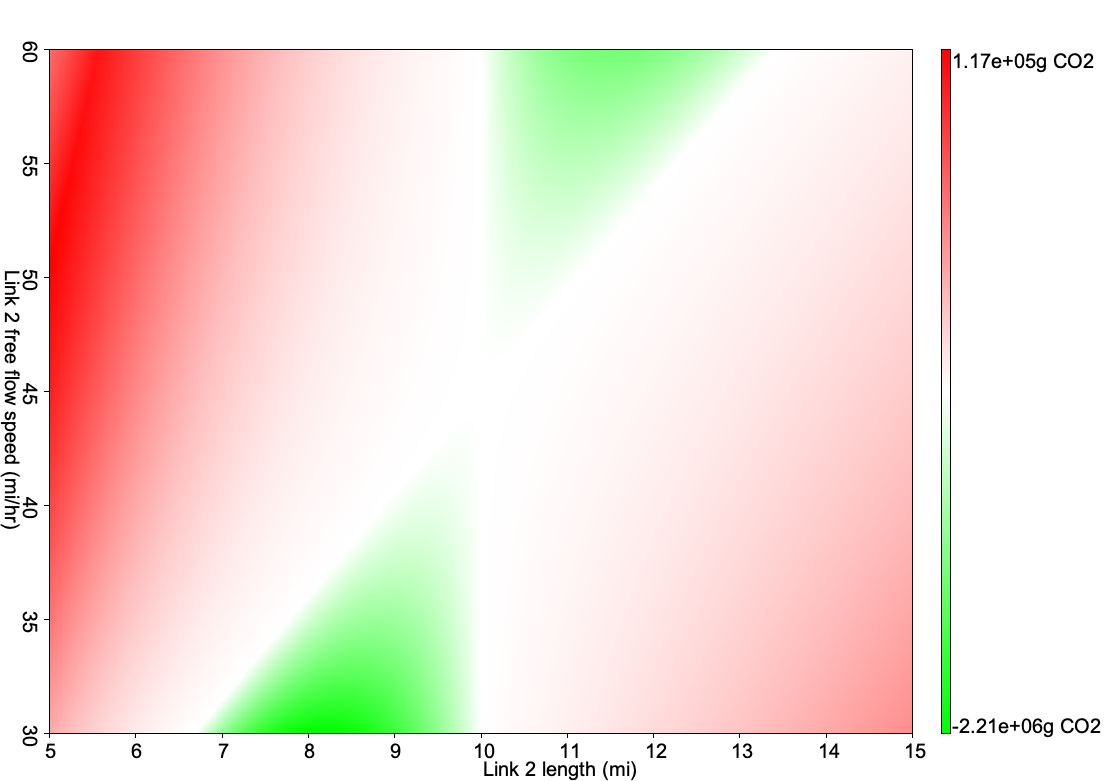}
        \caption{CO2 emissions: $\mathrm{TSE}_{\mathrm{eco-routing}}-\mathrm{TSE}_{\mathrm{time-routing}}$}
    \end{subfigure}

    \begin{subfigure}{0.75\textwidth}
        \includegraphics[width=\textwidth]{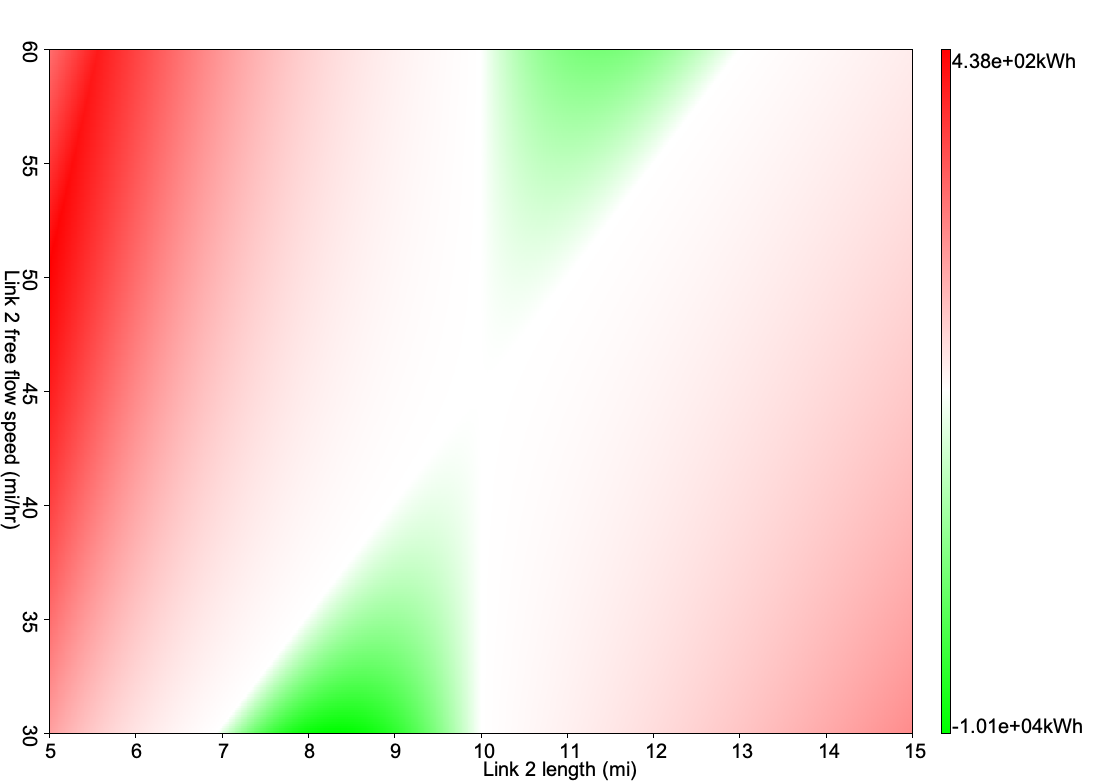}
        \caption{Fuel consumption: $\mathrm{TSFC}_{\mathrm{eco-routing}}-\mathrm{TSFC}_{\mathrm{time-routing}}$}
    \end{subfigure}
    \end{center}
    
    \caption{Change in emissions and fuel consumption caused by eco-routing to minimize fuel consumption with respect to changing the link length and free flow speed of link 2.}
    \label{2linkheatmap_fuel}
\end{figure}

It is not clear how common these results are. It might be necessary to choose a very specific set of parameters for eco-routing to increase emissions. 
We therefore keep the link 1 parameters constant while varying the parameters of link 2 to explore how the parameters affect the change in emissions and fuel consumption caused by eco-routing. We show the change in TSE and TSFC after switching 100\% of vehicles to eco-routing from time-routing, i.e.  $\mathrm{TSE}_{\mathrm{eco-routing}}-\mathrm{TSE}_{\mathrm{time-routing}}$ and $\mathrm{TSFC}_{\mathrm{eco-routing}}-\mathrm{TSFC}_{\mathrm{time-routing}}$. Negative values indicate that eco-routing improved TSE and TSFC, whereas positive values indicate that eco-routing made them worse. Figure \ref{2linkheatmap_emissions} shows these changes as a heatmap with respect to link 2 parameters for eco-routing to minimize emissions, and Figure \ref{2linkheatmap_fuel} shows heatmaps with eco-routing to minimize fuel consumption. 
Green indicates a reduction in TSE and TSFC, red indicates an increase, and the brightness of the colors indicates the magnitude, with white indicating zero or small changes.
The patterns are very similar with respect to the link 2 parameters, although the magnitude of the changes varies depending on the eco-routing strategy.

Overall, eco-routing reduces TSE and TSFC when link 2 is 0--30\% shorter than link 1 with a lower free flow speed, or when link 2 is 0--30\% longer than link 1 with a higher free flow speed. For many of the parameters chosen, the impact of eco-routing on TSE and TSFC is very small. However, when link 2 is 5--6mi in length, significant increases in TSE and TSFC from eco-routing are observed. The surprising result is that eco-routing reduces TSE and TSFC for a relatively small regime of link 2 parameters. In other words, given an uniform random selection of link 2 parameters $\ell_2\in [5,15]$ and $u^{\mathrm{ff}}_2\in [30,60]$, there is a 16.8\%  that eco-routing will reduce TSE, and a 83.1\% probability that eco-routing will make TSE worse. These results assume the capacities of link 1 and link 2 remain constant. Nevertheless, these results demonstrate that routing vehicles to minimize their own emissions can easily lead to increases in TSE and TSFC.

\subsection{City networks}

We also solved traffic assignment on city network data taken from Ben Stabler’s Transportation Networks GitHub page. For each city we had the following information about the network and the trips: the number of nodes and links, as well as the attributes corresponding to each link, and the total flow between all the possible combinations of origins and destinations within the city. In Table \ref{fig:citysizes} we can observe the sizes of the cities that were analyzed.

\begin{table}[]
\caption{Number of zones, nodes, and links for each of the cities that were analyzed.}
    \label{fig:citysizes}
\vspace{0.5cm}
\centering
\begin{tabular}{c|c|c|c|c|}
\cline{2-5}
                                          & Zones & Nodes & Links & Demand \\ \hline
\multicolumn{1}{|c|}{Barcelona}   & $110$     & $1020$       & $2522$  & $184679.561$                 \\ 
\multicolumn{1}{|c|}{Chicago}   & $387$    & $933$     & $2950$    & $1260907.44$             \\ 
\multicolumn{1}{|c|}{Eastern Massachusetts} & $74$    & $74$      & $258$    & $65576.3754$              \\
\multicolumn{1}{|c|}{Winnipeg}   & $147$    & $1052$     & $2836$      & $64784$           \\ \hline
\end{tabular}

\end{table}

\begin{figure}
    \centering
    \includegraphics[width=0.6\textwidth]{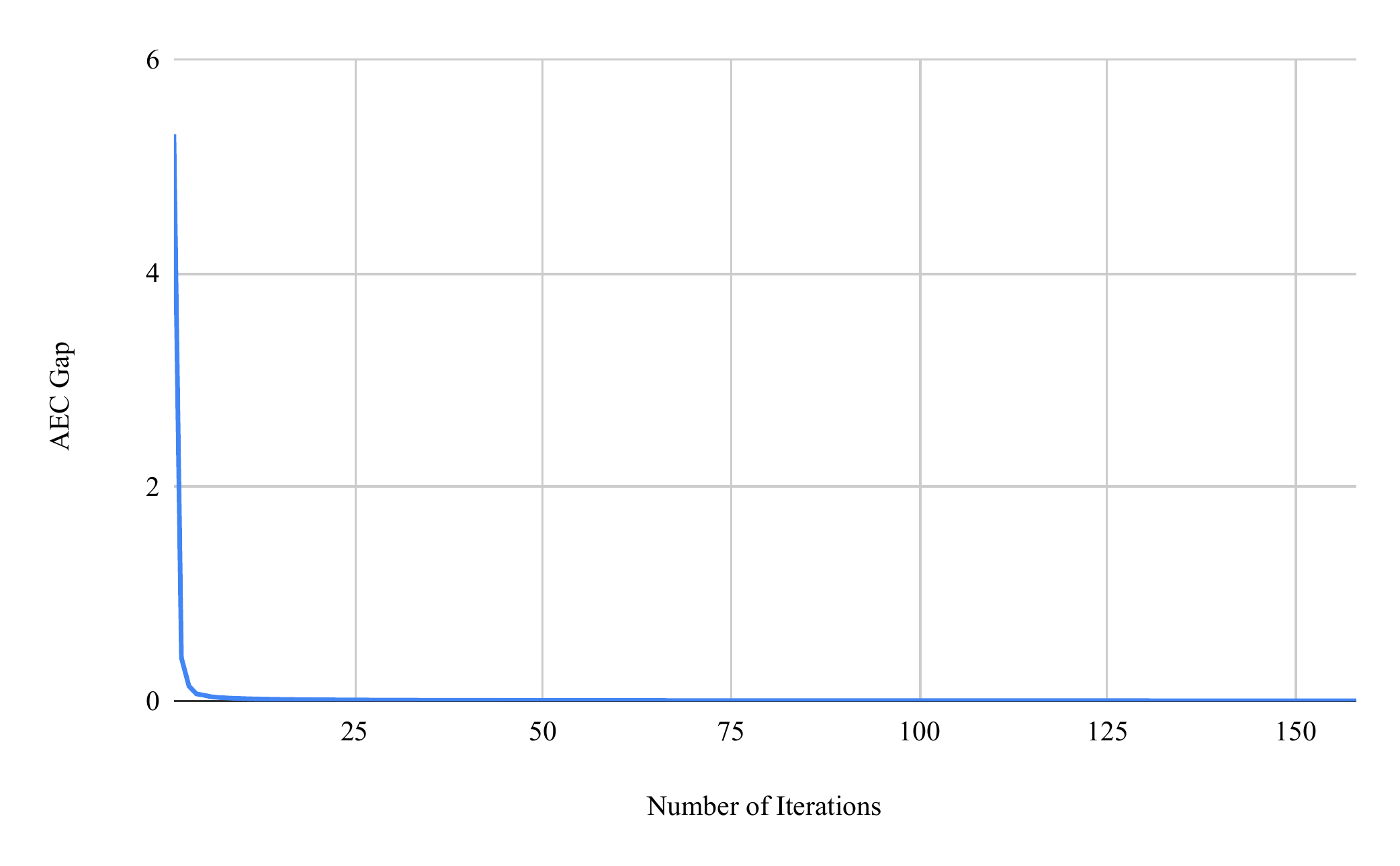}
    \caption{AEC Gap converges to zero as the equilibrium point is reached.}
    \label{fig:gap}
\end{figure}

In order to observe how implementing different levels of eco-routing demand would affect the total fuel consumption and total number of carbon emissions in these cities, we found the equilibrium link flows for multiple combinations of eco-routing and travel time demand, and then we obtained the total values for fuel consumption and emissions for the tested network. This equilibrium point was calculated both with respect to the fuel and the carbon emissions, using the values $\epsilon = 10^{-6}$ and $\epsilon = 10^{-4}$ respectively, with the exception of the emissions eco-routing analysis for Eastern Massachusetts that used $\epsilon = 10^{-8}$.  We can check that the equilibrium is reached in Figure \ref{fig:gap}. This was done for a range of eco-routing demand percentages from 0-100\% with a step size of 2\%.

\begin{figure}
     \centering
     \begin{subfigure}[b]{0.475\textwidth}
         \centering
         \includegraphics[width=\textwidth]{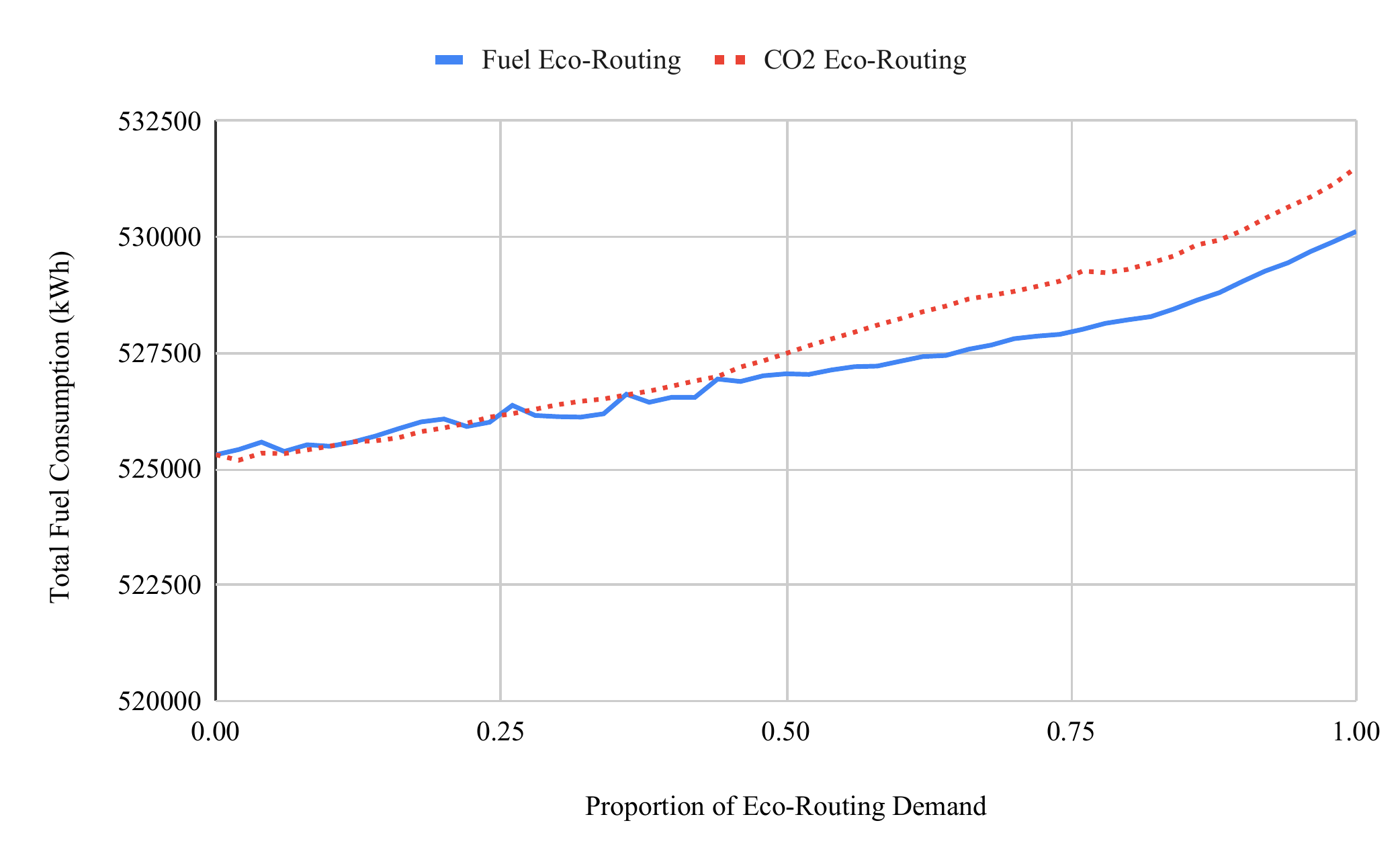}
         \caption{Total Fuel Consumption for Fuel and CO2 Eco-Routing}
         \label{fig:fuel_bcn}
     \end{subfigure}
     \hfill
     \begin{subfigure}[b]{0.475\textwidth}
         \centering
         \includegraphics[width=\textwidth]{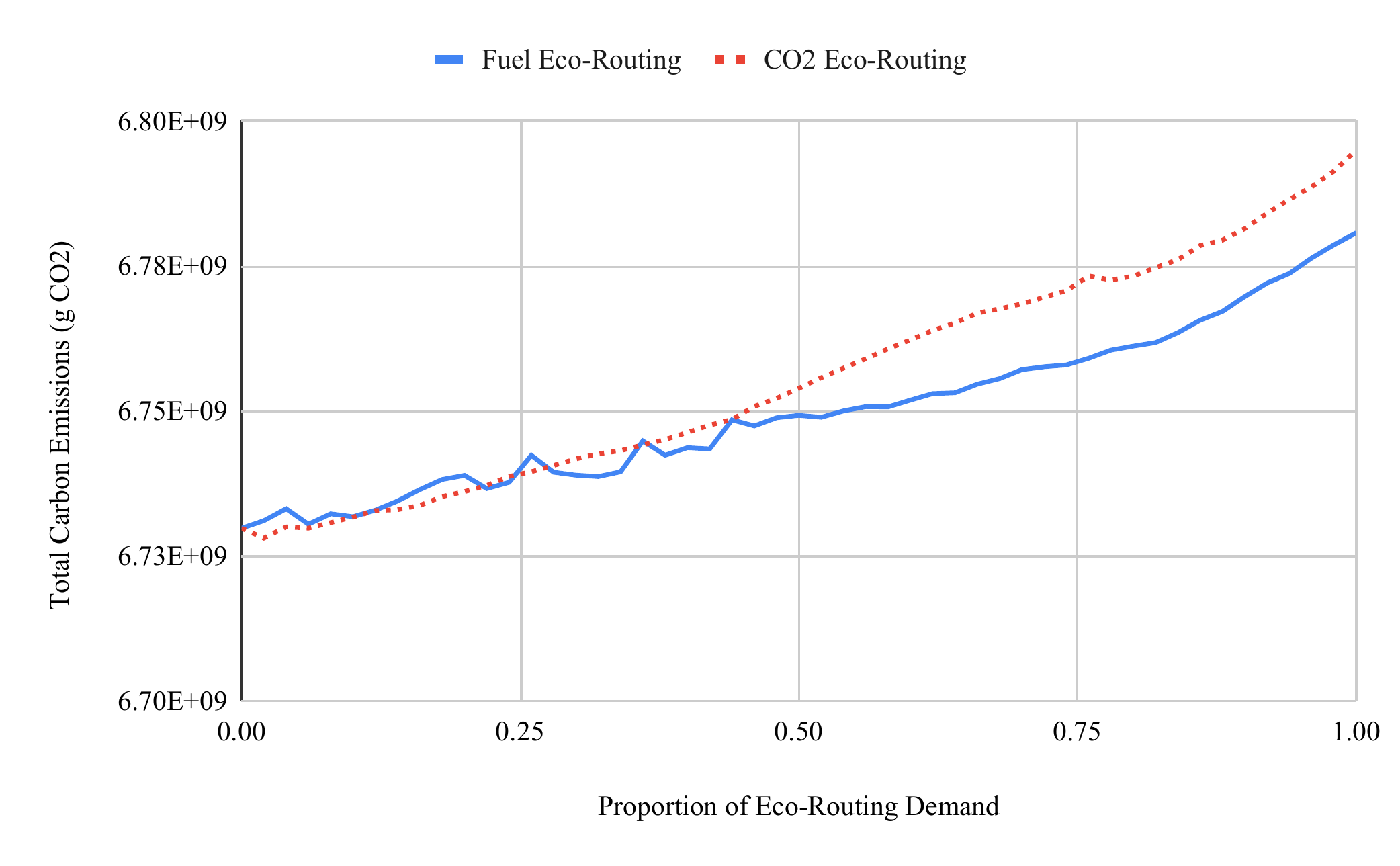}
         \caption{Total Carbon Emissions for Fuel and CO2 Eco-Routing}
         \label{fig:co2_bcn}
     \end{subfigure}
        \caption{Analysis of the total fuel consumption and total carbon emissions for Barcelona, Spain.}
        \label{fig:barcelona}
\end{figure}

\begin{figure}
     \centering
     \begin{subfigure}[b]{0.475\textwidth}
         \centering
         \includegraphics[width=\textwidth]{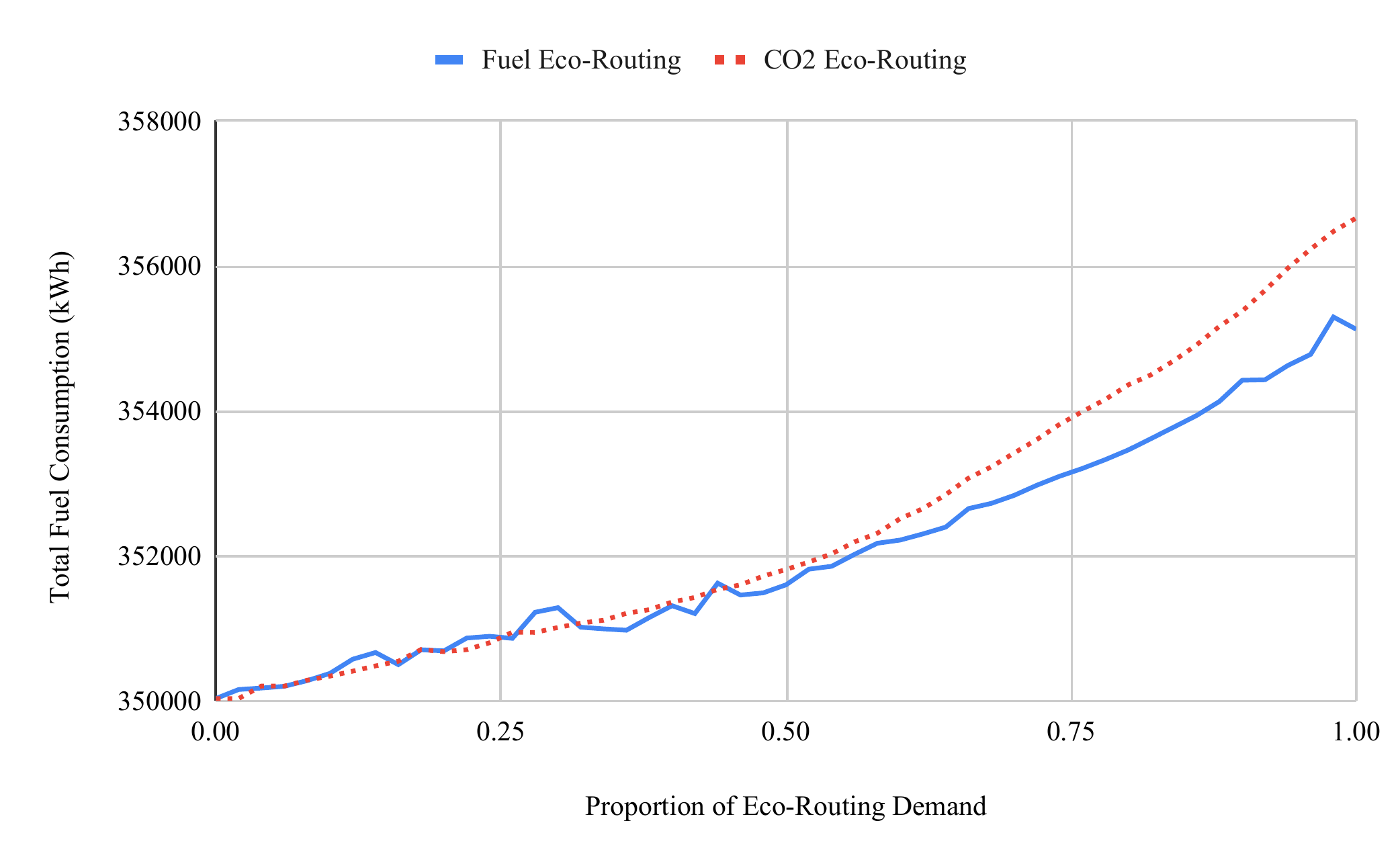}
         \caption{Total Fuel Consumption for Fuel and CO2 Eco-Routing}
         \label{fig:fuel_winn}
     \end{subfigure}
     \hfill
     \begin{subfigure}[b]{0.475\textwidth}
         \centering
         \includegraphics[width=\textwidth]{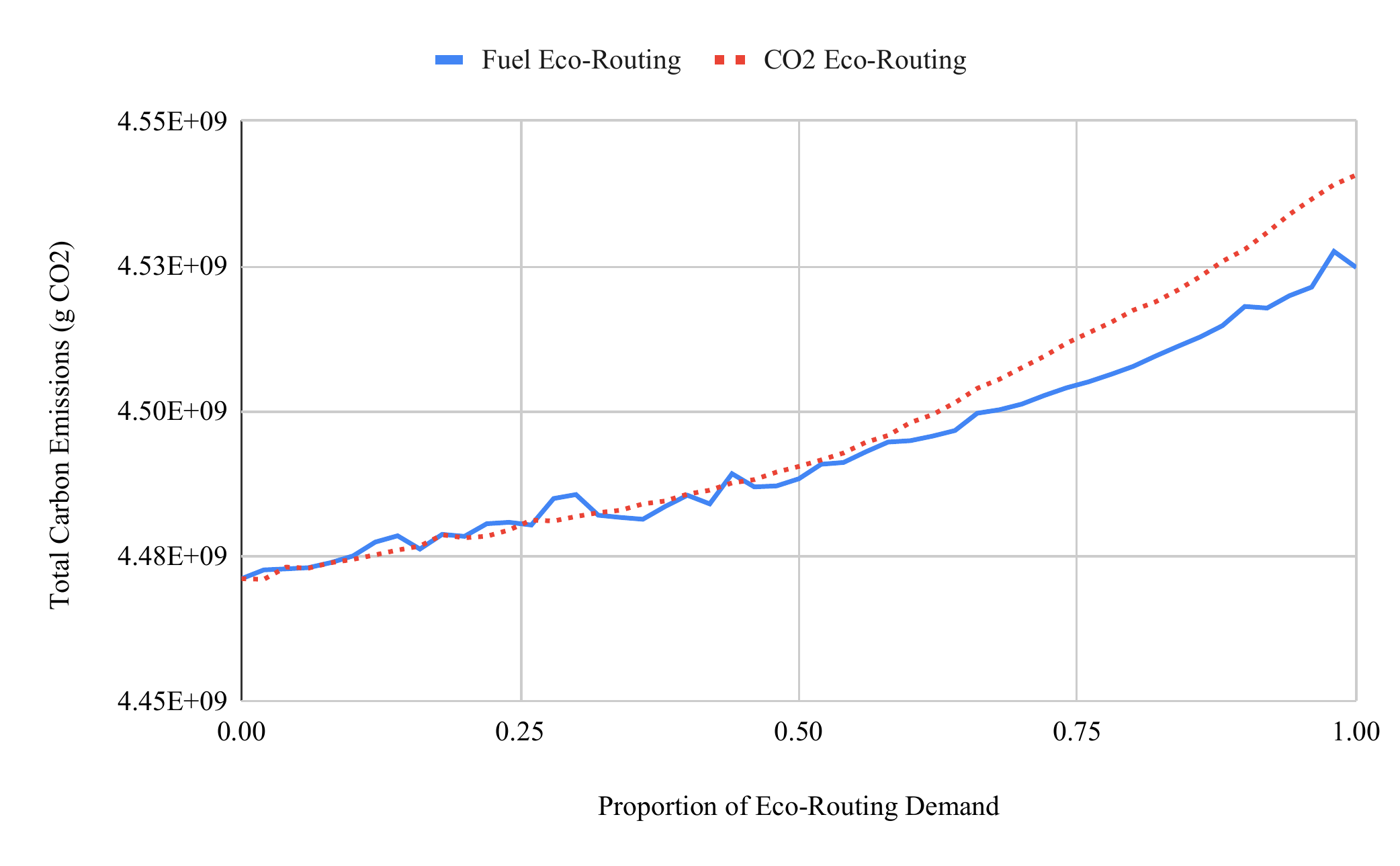}
         \caption{Total Carbon Emissions for Fuel and CO2 Eco-Routing}
         \label{fig:co2_winn}
     \end{subfigure}
        \caption{Analysis of the total fuel consumption and total carbon emissions for Winnipeg, Canada.}
        \label{fig:winnipeg}
\end{figure}

\begin{figure}
     \centering
     \begin{subfigure}[b]{0.475\textwidth}
         \centering
         \includegraphics[width=\textwidth]{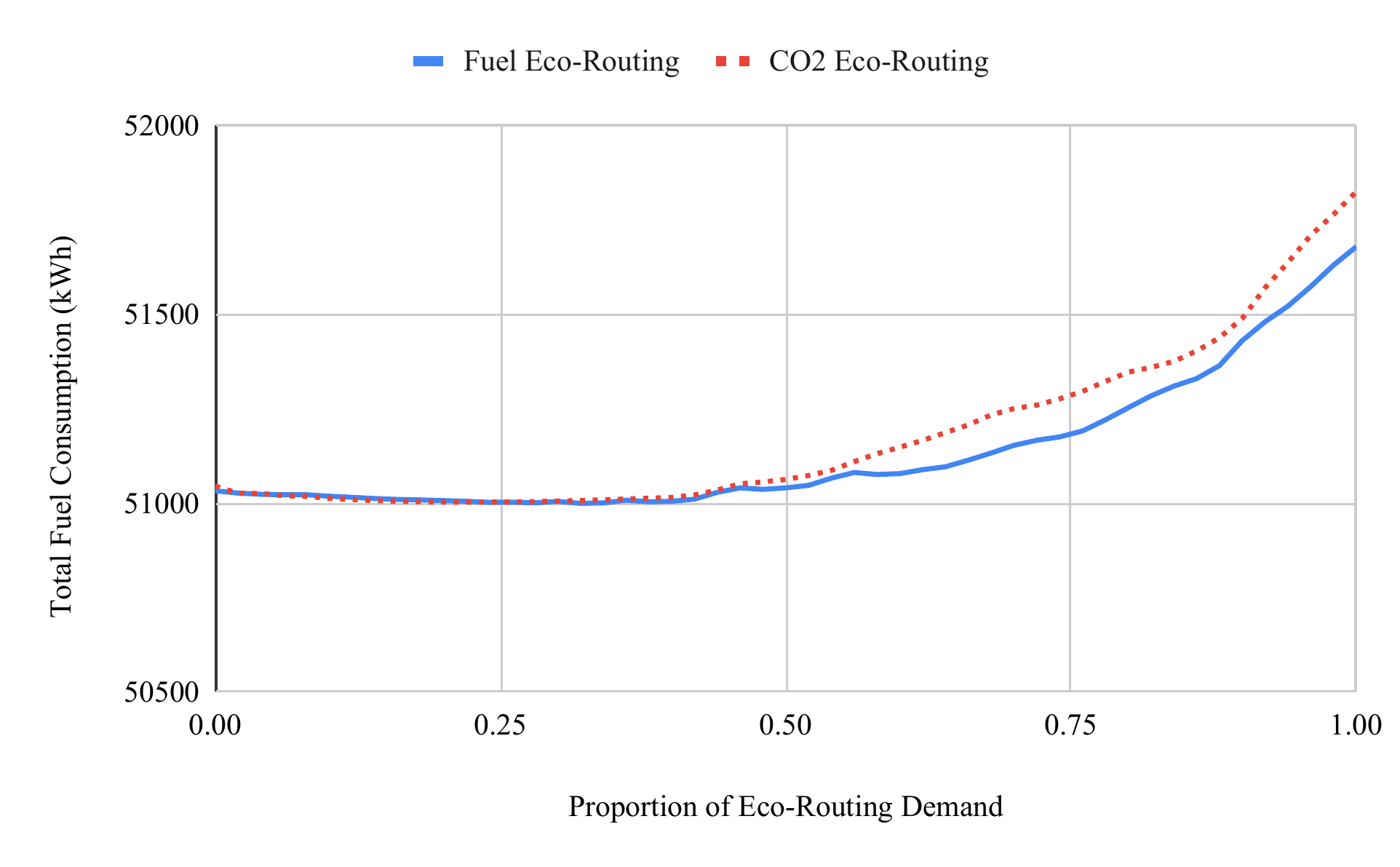}
         \caption{Total Fuel Consumption for Fuel and CO2 Eco-Routing}
         \label{fig:fuel_mass}
     \end{subfigure}
     \hfill
     \begin{subfigure}[b]{0.475\textwidth}
         \centering
         \includegraphics[width=\textwidth]{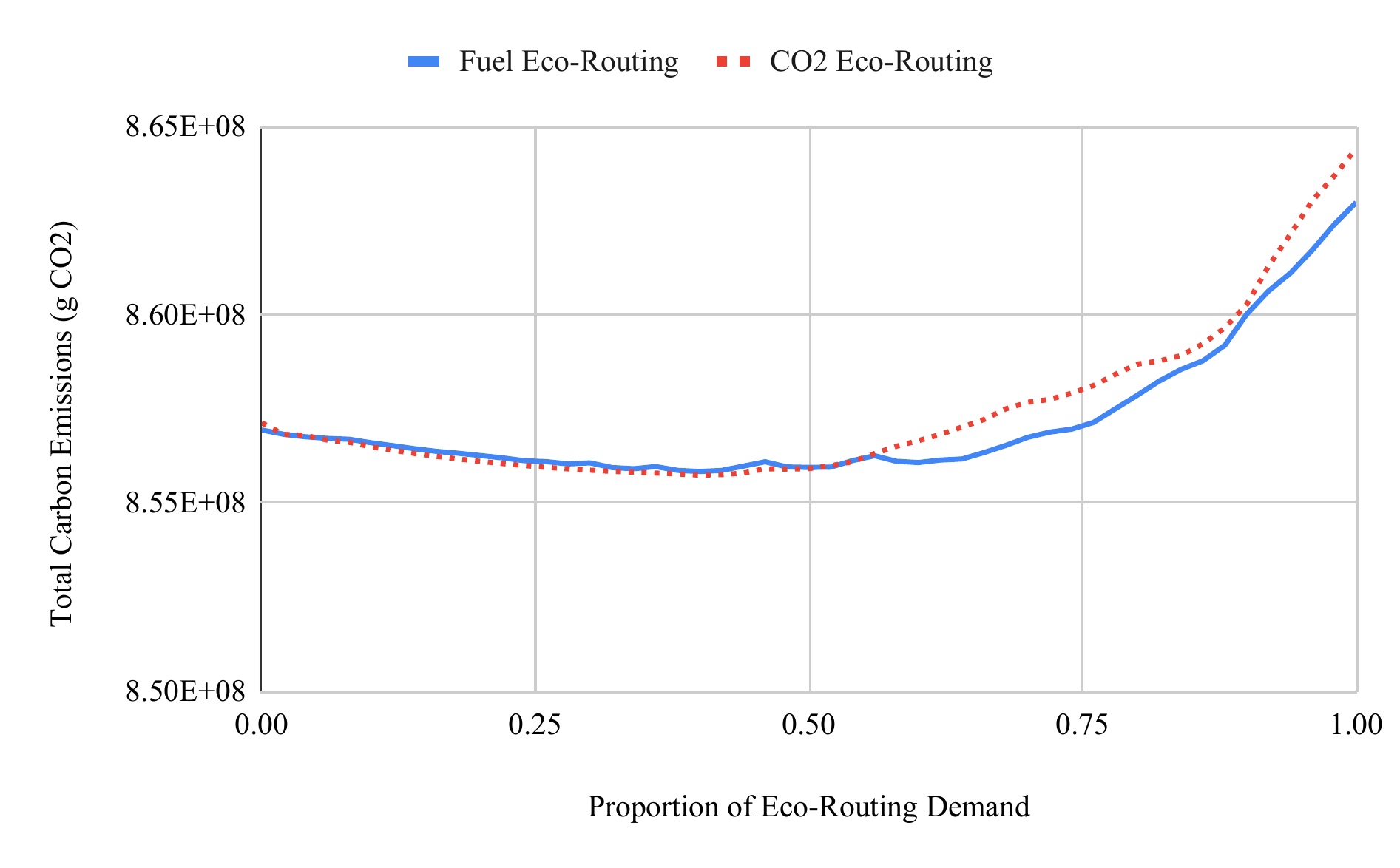}
         \caption{Total Carbon Emissions for Fuel and CO2 Eco-Routing}
         \label{fig:co2_mass}
     \end{subfigure}
        \caption{Analysis of the total fuel consumption and total carbon emissions for Eastern Massachusetts.}
        \label{fig:massachusetts}
\end{figure}

\begin{figure}
     \centering
     \begin{subfigure}[b]{0.475\textwidth}
         \centering
         \includegraphics[width=\textwidth]{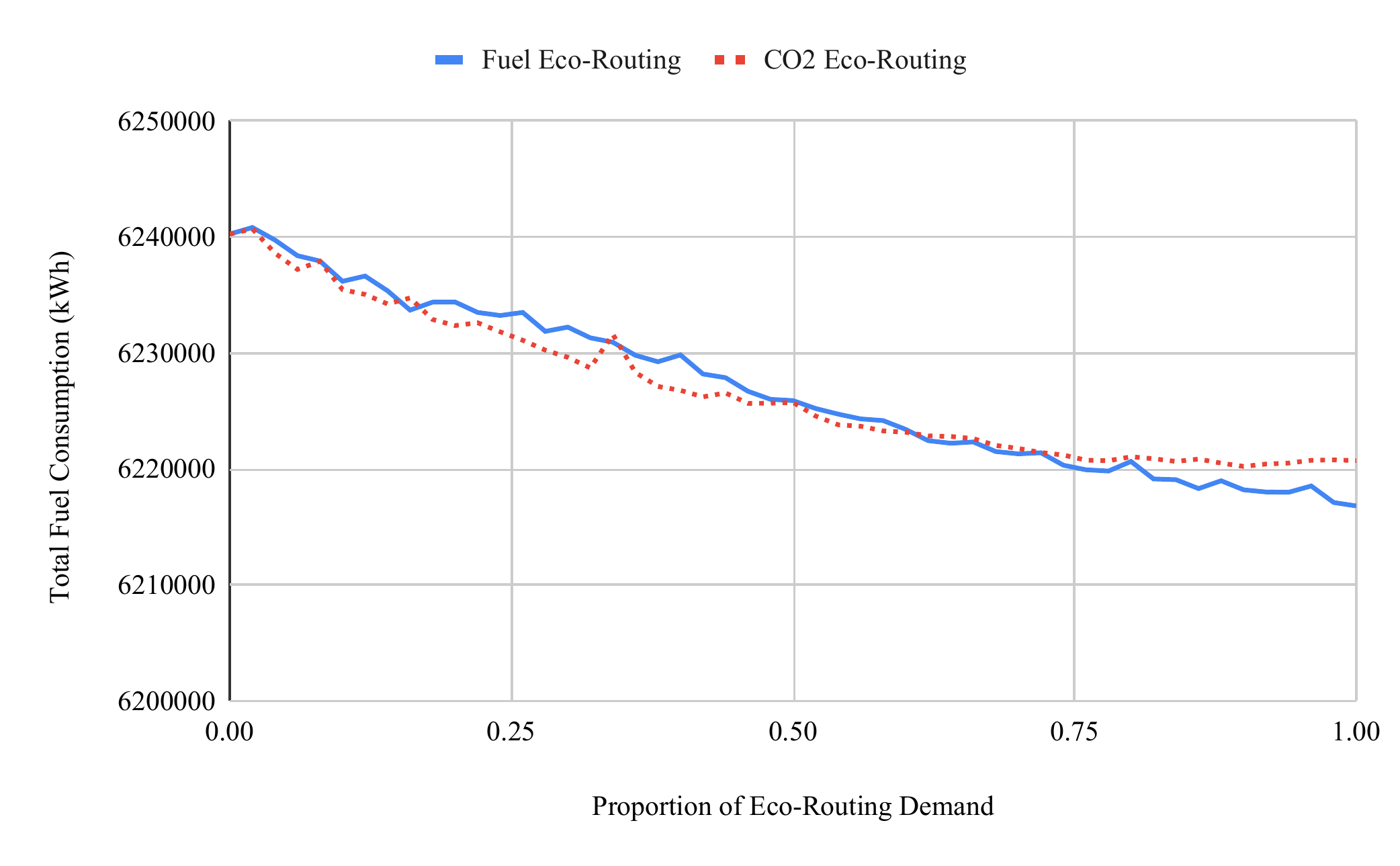}
         \caption{Total Fuel Consumption for Fuel and CO2 Eco-Routing}
         \label{fig:fuel_chi}
     \end{subfigure}
     \hfill
     \begin{subfigure}[b]{0.475\textwidth}
         \centering
         \includegraphics[width=\textwidth]{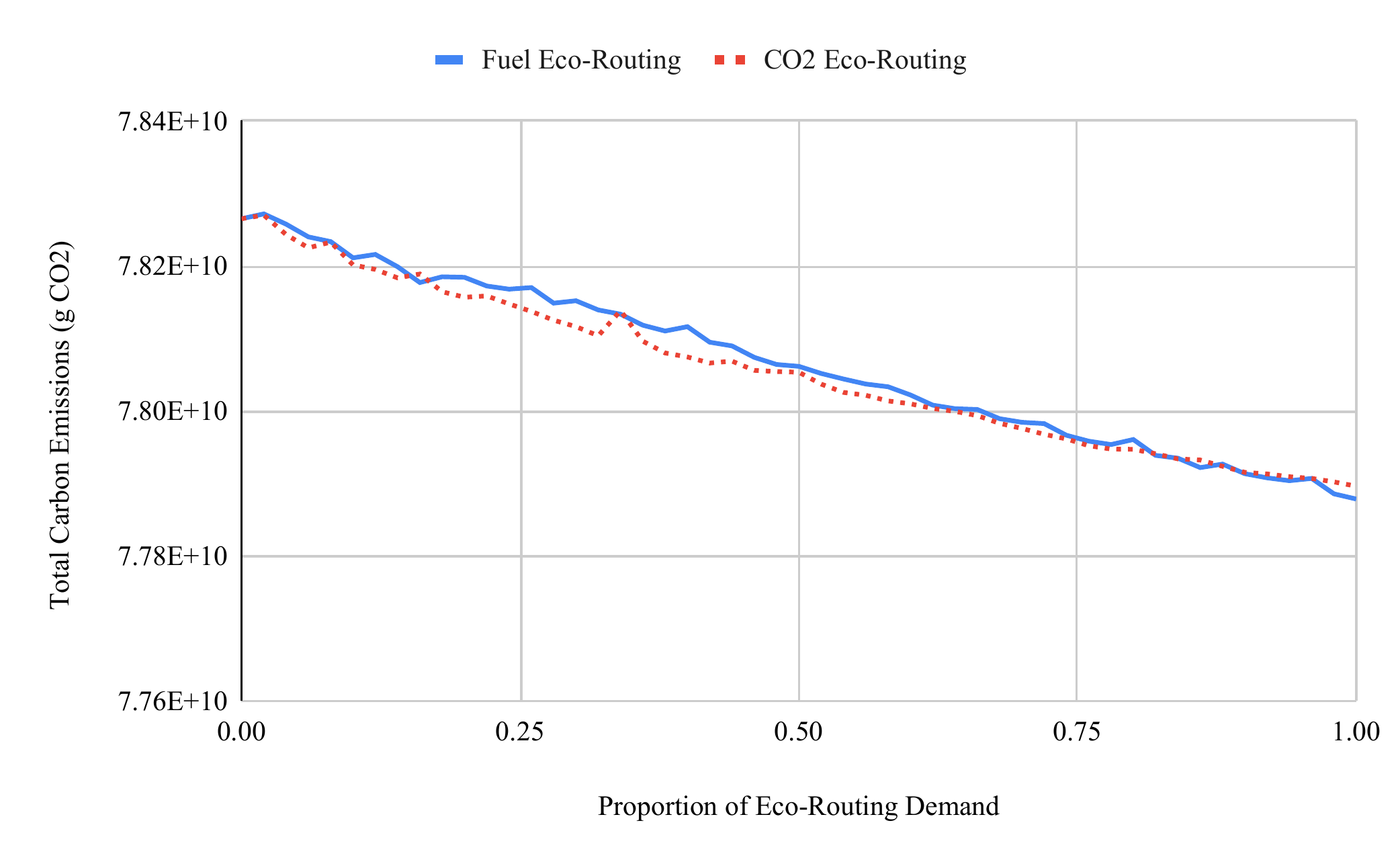}
         \caption{Total Carbon Emissions for Fuel and CO2 Eco-Routing}
         \label{fig:co2_chi}
     \end{subfigure}
        \caption{Analysis of the total fuel consumption and total carbon emissions for Chicago, Illinois.}
        \label{fig:chicago}
\end{figure}

As we can see in Figure \ref{fig:massachusetts}, the total fuel consumption and total carbon emissions for both the fuel-routing and CO2 emissions-routing in Eastern Massachusetts followed a really similar growth pattern. For low levels of eco-routing demand, the TSFC and TSE have a decreasing trend but after hitting a minimum, they increase for the rest of the proportions of eco-routing demand. For the cities of Barcelona and Winnipeg, as we can observe in Figures \ref{fig:barcelona} and \ref{fig:winnipeg}, we found that the fuel-routing and emissions-routing analysis had an increasing trend for both the TSFC and TSE. For all these cities, the CO2 emissions eco-routing and fuel eco-routing results increased at a very close rate until about 50\% of demand, where the emissions-routing increased at a higher rate. We also found a case where the TSFC and TSE decreased when applying both fuel and CO2 eco-routing. We can see this in Figure \ref{fig:chicago}, where the total fuel consumption and the total carbon emissions for Chicago had a decreasing trend all the way from 0\% to 100\% of eco-routing demand. We notice that the cities that resulted in an increase of total emissions and total fuel consumption were significantly smaller compared to Chicago, in which we observed a decrease of both and was the largest city analyzed. This suggests that the eco-routing model is more efficient in large cities, rather than in small ones.

The results from analyzing cities' networks showed that for higher levels of eco-routing demand, the total fuel consumption and total emissions increased for most of the analyzed cities (Barcelona, Winnipeg, Eastern Massachusetts), while for one of them, it decreased (Chicago). It is important to note that the graphs for both total fuel and total emissions calculated with the same eco-routing parameter in a city had the same shape, which indicates that there was a clear direct relationship between them.

Intuitively, it is thought that a high level of eco-routing would result in a decrease of TSFC and TSE, as we can observe happens in Chicago. However, the results obtained for cities like Barcelona, Winnipeg, and Eastern Massachusetts demonstrate that, paradoxically, it is possible for them to increase with the demand. It would make sense to believe that the reasoning behind is the same pattern that was found in a 2-link network, implying that most of the relations between the links that form the cities of Barcelona, Eastern Massachusetts, and Winnipeg have the characteristics that cause eco-routing to increase generally.

\section{Conclusions}

Drivers using Google Maps' default routing behavior are now being routed to minimize their own carbon emissions~\citep{GoogleMaps}, and this paper demonstrates that such routing behavior could actually increase total system emissions (TSE) and total system fuel consumption (TSFC). 
This paper defined a multiclass user equilibrium where some vehicles seek to minimize their own travel time, and other vehicles seek to minimize their own CO2 emissions or fuel consumption (eco-routing), and explored the user equilibrium behavior in terms of TSE and TSFC. We demonstrate that it is very possible for eco-routing to increase TSE and TSFC both on a simple 2-link network and on several city networks. 
Since individual vehicles seek to minimize their own emissions or fuel consumption, the changes in traffic flow patterns and congestion can lead to more vehicles experiencing higher emissions and higher overall TSE. 
These results are consistent with well-known characteristics of user equilibrium behavior such as the \citet{braess1968paradoxon} paradox, but have not yet been demonstrated for eco-routing and are particularly timely given the recent shift in implementation of eco-routing by Google Maps~\citep{GoogleMaps}.

We acknowledge several limitations of this study. 
We use a non-convex multiclass traffic assignment model for the city network evaluations, which could have multiple equilibrium solutions. Consequently, although we observed equilibria in which eco-routing increases TSE and TSFC, other equilibria with different impacts on TSE and TSFC could exist. However, this is not a limitation of our 2-link network results because we did not include multiclass flow in those.
Estimating emissions is far more complex than using a simple regression model from \citet{gardner2013framework}. 
A more realistic model by \citet{ahn2013network} found that eco-routing reduced TSFC in 2 specific networks, but that does not contradict our results. Eco-routing can be beneficial in some city networks while being harmful in others.  
Traffic congestion also varies with time, which is not captured by the static traffic assignment model. However, we note that the \citet{braess1968paradoxon} paradox of time-routing has been demonstrated in dynamic traffic assignment~\citep{zhang2008braess}, and more realistic time-dependent traffic models have created new opportunities for time-routing to increase total system travel time after network improvements~\citep{daganzo1998queue}. Therefore, we believe that the increases in TSE and TSFC observed in our static traffic assignment model are likely to occur in some more realistic models and scenarios as well.

We hope that awareness of the potential limitations of eco-routing will inspire better methods of eco-routing to reduce TSE and/or TSFC. If static traffic network routes are used, finding the system optimal solution that minimizes TSE is straightforward. However, the assumptions of static traffic assignment are unrealistic. Although methods exist to solve system optimal dynamic traffic assignment~\citep{ziliaskopoulos2000linear, li2003decomposition}, it is far more computationally difficult and cannot yet be solved to optimality on large city networks. Nevertheless, heuristics such as dynamic tolling exist to reduce the total system travel time in mesoscopic and microscopic simulation models~\citep[e.g.][]{sharon2017network}. Similar heuristics might be developed to minimize TSE. If such routing heuristics could be implemented into commonly-used navigation apps, they might achieve an overall decrease in greenhouse gas emissions from transportation.

%
%
\bibliography{ascexmpl-new}

\end{document}